\documentclass[aps,showpacs,twocolumn]{revtex4}
\usepackage{epsfig}

\begin{document}

\title{The structure of pentaquarks $\Omega_c^0$ in the chiral quark model}

\author{Gang Yang, Jialun Ping{\footnote{jlping@njnu.edu.cn, corresponding author}}}

\affiliation{$^1$Department of Physics and Jiangsu Key Laboratory for Numerical
Simulation of Large Scale Complex Systems, Nanjing Normal University, Nanjing 210023, P. R. China}

\begin{abstract}
  Recently, the experimental results of LHCb Collaboration suggested the existence of five new
  excited states of $\Omega_c^0$, $\Omega_c(3000)^0$, $\Omega_c(3050)^0$, $\Omega_c(3066)^0$,
  $\Omega_c(3090)^0$ and $\Omega_c(3119)^0$, the quantum numbers of these new particles are not
  determined now. To understand the nature of the
  states, a dynamical calculation of 5-quark systems with quantum numbers $IJ^P=0(\frac{1}{2})^-$,
  $0(\frac{3}{2})^-$ and $0(\frac{5}{2})^-$ is performed in the framework of chiral quark model with
  the help of gaussian expansion method. The results show the $\Xi\bar{D}$, $\Xi_c\bar{K}$
  and $\Xi_c^*\bar{K}$ are possible the candidates of these new particles. The distances between
  quark pairs suggest that the nature of pentaquark states.
\end{abstract}


\maketitle

\section{Introduction}

Recently, CERN announced an exceptional new discovery that was made by the LHCb, which unveiled five
new states all at once~\cite{lhcb}. Each of the five particles were found to be the excited states of $\Omega_c^0$,
a particle with three quarks, $css$. These particle states are named, according to the standard convention,
$\Omega_c(3000)^0$, $\Omega_c(3050)^0$, $\Omega_c(3066)^0$, $\Omega_c(3090)^0$ and $\Omega_c(3119)^0$.
Just after the announcement, the theoretical interpretations were proposed. S. S. Agaev {\em et al.} interpreted two of
these excited charmed baryons ($\Omega_c(3066)^0$ and $\Omega_c(3119)^0$) as the first radial excitation with
($2S, 1/2^+$) and ($2S,3/2^+$), respectively in QCD sum rules~\cite{SSAKA}. The same conclusion is proposed by
H. X. Chen {\em et al.}~\cite{HXCSLZ} in studying the decay properties of $P$-wave charmed baryons from light-cone
QCD sum rules, besides they also suggest that one of these $\Omega_c^0$ states $(\Omega_c(3000)^0,
\Omega_c(3050)^0 or~\Omega_c(3066)^0)$ as a $J^P=1/2^-$ state, the rest two states is with $J^P=3/2^-$ and $J^P=5/2^-$.
In Ref.~\cite{MKJLR}, Karliner and Rosner suggested that the parity was negative for all of the five states,
two $J^P=1/2^-$ states ($\Omega_c(3000)^0$ and $\Omega_c(3050)^0$), two $J^P=3/2^-$ states ($\Omega_c(3066)^0$ and
$\Omega_c(3090)^0$), and the last one is $\Omega_c(3119)^0$ $J^P=5/2^-$.
These exciting announcements and the theoretical work along with the pentaquarks $P_c^+$ discovered also
by the LHCb Collaboration in 2015~\cite{M6} have bring us lots of peculiar understanding to the world of
microcosmic particles.

The quantum numbers of these new particles are not determined for the moment, and the explanation of
them as the excited states of $q^3$ baryon is reasonable. However, the possibility of the multi-quark
candidates of these excited states cannot be excluded.
The ground states of $\Omega_c$ have been observed experimentally, $\Omega_c(2695)^0$ with $J^P=\frac{1}{2}^+$
and $\Omega_c(2770)^0$ with $J^P=\frac{3}{2}^+$. The excited energies of the newly reported states with
respect to the ground states are 230-424 MeV, which are enough to excite light quark-antiquark pair from the
vacuum. From the masses of $\Xi_c$ baryon and $K$ meson, 2468 MeV and 495 MeV, we have the threshold for
$\Xi_c-\bar{K}$ state around 2963 MeV. It is expected that the 5-$q$ components will play a role in these
$\Omega_c$'s. In Ref.~\cite{OmegaS}, spectrum of low-lying pentaquark states with strangeness $S=-3$ and
negative parity is studied in three kinds of constituent quark models. The results indicate that the lowest
energy state $\Omega^*$ is around 1.8 GeV, which is about 200 MeV lower than predictions of various quenched
three quarks models, and the energy cost to excite ground state of $\Omega$ to a 5-quark state is less than
that to an orbital excitation.

The interesting in pentaquark is revived after the observation of the exotic hadrons, $P_c^+(4380)$ and
$P_c^+(4450)$ in the decay of $\Lambda_b^0$, $\Lambda_b^0 \rightarrow J/\psi K^-p$ by the LHCb Collaboration
lately~\cite{M6}, there are a lots of theoretical calculations have been performed to investigate these two
exotic states~\cite{YG,huang,RChen,JHe,HXChen,ZGWang,Roca,ZQ,ATS,PRD92,review}, even though the $\Theta^+(1540)$
pentaquark was reported by several experimental groups~\cite{M1,M2,M3} in 2003 and has been denied by JLab with
more higher precision results~\cite{notheta} (LEPS Collaboration still insisted on the existence of pentaquark
$\Theta^+(1540)$~\cite{theta2}). Besides, it is shown that there should be notable five-quark components in
the baryon resonances~\cite{PRC80065, PRC82062, EPJA39195}. In addition, the valence-sea quark mixing
(Fock space expansion) model (${q^3}$+ ${q^3}{q}\bar{q}$) of nucleon ground state had been used to explain
the mysterious proton spin structure well~\cite{wang}. Such a sea quark excitation model had also been used to
show that the ${q^3}q\bar{q}$ excitation is more favorable than the $p$-wave excitation in $q^3$ configuration
for $1/2^-$ baryons~\cite{M4}.

Quark model is the most common approach to multi-quark system. With the recent experimental data on multi-quark
states and the development of quark model, It is expected to perform a serious calculation of multi-quark
state in the framework of quark model. In the present work, the chiral quark model (ChQM) is employed to study
the pentaquark states $\Omega^0_c$. To find the structure of the pentaquark states, a general, powerful
method of few-body system, gaussian expansion method (GEM)~\cite{GEM} is used to do the calculation.
The GEM has been successfully applied to many few-body systems, light nuclei, hyper-nuclei, hadron physics
and so on~\cite{GEM}. It suits for both of compact multi-quark systems and loosely bound molecular states.
In this approach, the four relative orbital motions of the system are expanded by gaussians with various
widths. By taking into account of all the possible couplings for color-flavor-spin degrees of freedom,
the structure of the system determined by its dynamics can be found.

The structure of the paper is as follows. In section II the quark model, wave-functions and calculation
method is presented. Section III is devoted to the calculated results and discussions. A brief summary
is given in the last section.

\section{model and wave function}

The chiral quark model has achieved a success both in describing the hadron spectra and hadron-hadron
interaction. In this model, the constituent quark and antiquark interact with each other through the
Goldstone boson exchange and the effective one-gluon-exchange, in addition to the phenomenological color
confinement. Besides, the scalar nonet (the extension of chiral partner $\sigma$-meson) exchange are also
introduced. The details of the model can be found in Ref.\cite{ChQM}. So the Hamiltonian in the present
calculation takes the form
\begin{widetext}
\begin{eqnarray}
H & = & \sum_{i=1}^{n}\left( m_i+\frac{p^2_i}{2m_i}\right)-T_{CM}
      + \sum_{j>i=1}^{n}\left[ V_{CON}({{\bf r}_{ij}})+V_{OGE}({{\bf r}_{ij}})
      +V_{\chi}({{\bf r}_{ij}})+V_{s}({{\bf r}_{ij}})\right] , \\
V_{CON}({{\bf r}_{ij}}) & = & \mbox{\boldmath $\lambda$}_i^c\cdot\mbox{\boldmath $\lambda$}_j^c
   \left[-a_c (1-e^{-\mu_cr_{ij}})+\Delta \right] , \nonumber \\
V_{OGE}({{\bf r}_{ij}}) & = & \frac{1}{4}\alpha_s\mbox{\boldmath $\lambda$}_i^c\cdot\mbox{\boldmath $\lambda$}_j^c
   \left[ \frac{1}{r_{ij}}-\frac{1}{6m_im_j}\mbox{\boldmath $\sigma$}_i\cdot\mbox{\boldmath $\sigma$}_j
   \frac{e^{-r_{ij}/r_0(\mu)}}{r_{ij}r^2_0(\mu)}\right] , ~~~r_0(\mu)=\hat{r}_0/\mu, \nonumber \\
V_{\chi}({{\bf r}_{ij}}) & = & v_{\pi}({{\bf r}_{ij}})
   \sum_{a=1}^{3}\mbox{\boldmath $\lambda$}_i^a\cdot\mbox{\boldmath $\lambda$}_j^a
   +v_{K}({{\bf r}_{ij}})\sum_{a=4}^{7}\mbox{\boldmath $\lambda$}_i^a\cdot\mbox{\boldmath $\lambda$}_j^a
   +v_{\eta}({{\bf r}_{ij}})[\mbox{\boldmath $\lambda$}_i^8
      \cdot\mbox{\boldmath $\lambda$}_j^8\cos\theta_{P}-\mbox{\boldmath $\lambda$}_i^0
      \cdot\mbox{\boldmath $\lambda$}_j^0\sin\theta_{P}],   \\
v_{\chi}({{\bf r}_{ij}}) & = & \frac{g^2_{ch}}{4\pi}\frac{m^2_\chi}{12m_im_j}
	\frac{\Lambda^2_\chi}{\Lambda^2_\chi-m^2_\chi}m_\chi
	\left[ Y(m_{\chi}r_{ij})-\frac{\Lambda^3_\chi}{m^3_\chi}Y (\Lambda_{\chi}r_{ij}) \right]
	\mbox{\boldmath $\sigma$}_i\cdot\mbox{\boldmath $\sigma$}_j,    ~~~~\chi=\pi, K,\eta, \nonumber \\
V_{s}({{\bf r}_{ij}}) & = & v_{\sigma}({{\bf r}_{ij}})(\mbox{\boldmath $\lambda$}_i^0
\cdot\mbox{\boldmath $\lambda$}_j^0)+v_{a_0}({{\bf r}_{ij}})
   \sum_{a=1}^{3}\mbox{\boldmath $\lambda$}_i^a\cdot\mbox{\boldmath $\lambda$}_j^a
   +v_{\kappa}({{\bf r}_{ij}})\sum_{a=4}^{7}\mbox{\boldmath $\lambda$}_i^a\cdot\mbox{\boldmath $\lambda$}_j^a
   +v_{f_0}({{\bf r}_{ij}})(\mbox{\boldmath $\lambda$}_i^8\cdot\mbox{\boldmath $\lambda$}_j^8),   \\
v_{s}({{\bf r}_{ij}}) & = & -\frac{g^2_{ch}}{4\pi} \frac{\Lambda^2_s}{\Lambda^2_s-m^2_s}m_s
	\left[ Y(m_{s}r_{ij})-\frac{\Lambda_s}{m_s}Y(\Lambda_{s}r_{ij})\right], ~~~~s=\sigma, a_0,\kappa ,f_0.
  \nonumber
\end{eqnarray}
\end{widetext}
All the symbols take their usual meanings. $\mu$ is the reduced mass of two interacting quarks.
To simplify the calculation, only the central parts of the interactions are employed in the present work
to consider the ground state of multi-quark system. The model parameters are fixed by fitting the spectrum
of baryons and mesons and their values are listed in Table I, the calculated masses of baryons and mesons
are shown in Table II. There are two sets of parameters are given, the fixed quark-gluon coupling constant
is used in the set I, the set II has the running coupling constants which are given as
$$
\alpha_s=\frac{\alpha_0}{\ln((\mu^2+\mu_0^2)/\Lambda_0^2)}.
$$
It is worth to mention that the above quark-quark interaction is assumed to be universal
according to the "Casimir scaling"~\cite{Bali}, it can be applied to the multi-quark system directly.
The possible multi-body interaction in the multiquark system is not considered, although it may give
different spectra of multiquark states~\cite{multibody}.
\begin{table}[ht]
\caption{Quark model parameters. The masses of mesons take their experimental values.
$m_{\pi}=0.7$ fm$^{-1}$, $m_{K}=2.51$ fm$^{-1}$, $m_{\pi}=2.77$ fm$^{-1}$.  \label{model}}
\begin{tabular}{c|cccc} \hline
                   &                    & set I & set II \\ \hline
                   &$m_u$=$m_d$ (MeV)   &313 & 313 \\
   Quark mass      &$m_s$ (MeV)  & 555 & 555 \\
                   &$m_c$ (MeV)  &1752 & 1752 \\   \hline
                   &$\Lambda_\pi=\Lambda_{\sigma}$ (fm$^{-1}$)  &4.20 & 4.20\\
  Goldstone boson  &$\Lambda_K=\Lambda_\eta$ (fm$^{-1}$)  &5.20 & 5.20\\
                   &$\theta_P(^\circ)$  &-15  & -15 \\
                   &$g^2_{ch}/(4\pi)$  &0.54 & 0.54 \\ \hline
       SU(3)       &  $m_{s}$ (fm$^{-1}$)      &4.97 & 4.97 \\
   Scalar nonet & $\Lambda_{s}$ (fm$^{-1}$)   &5.20 & 5.20 \\
  $s=\sigma, a_0,\kappa ,f_0$ & $m_{\sigma}$ (fm$^{-1}$)      &3.42 & 3.42 \\  \hline
            &$a_c$ (MeV)  &180 & 184.08 \\
Confinement &$\mu_c$ (fm$^{-1})$  &0.645 & 0.634 \\
            &$\Delta$ (MeV)  &55.5 & 40.249\\         \hline
       &     &   & $\alpha_0 =1.293$ \\
 OGE   &$\alpha_s~$  &0.69   & $\Lambda_0=1.5585$ fm$^{-1}$ \\
       &        &             & $\mu_0=621.5$ MeV \\
       &$\hat{r}_0~$(MeV~fm)  &28.170 & 43.882\\     \hline
\end{tabular}
\end{table}

\begin{table}[ht]
\caption{Masses of baryon and meson in ChQM (unit: MeV). \label{SBM}}
\begin{tabular}{ccccccc} \hline
   $P$ &$N(939)$ &$\Delta(1232)$  &$\Omega(1672)$ &$\Lambda(1116)$  &$\Sigma(1189)$ &$\Xi(1315)$\\
\hline
\multicolumn{7}{c}{set I} \\ \hline
  $+$ &936 &1208 &1643 &1154 &1173 & 1362 \\
  $-$ &1575 &1625 &2203 &1772 &1777 & 1981 \\ \hline
\multicolumn{7}{c}{set II} \\ \hline
  $+$ &939 &1231 &1671 &1187 &1209 & 1408 \\
  $-$ &1661 &1716 &2301 &1889 &1895 & 2098 \\ \hline
  & $\Sigma^*(1383)$ &$\Xi^*(1532)$  &$\Omega_c(2695)$ &$\Omega_c(2765)$ & $\Xi_c(2467)$ & $\Xi_c^*(2645)$ \\
\hline
\multicolumn{7}{c}{set I} \\ \hline
  $+$ &1342 &1488 &2675 &2748 &2541 & 2603 \\
  $-$ &1805 &1999 &3257 &3282 &3086 & 3093 \\ \hline
\multicolumn{7}{c}{set II} \\ \hline
 $+$ &1393 &1539 &2748 &2818 &2629  & 2727 \\
 $-$ &1928 &2119 &3378 &3389 &3145  & 3166 \\ \hline
\multicolumn{7}{c}{set I} \\ \hline
  $P$ &$\pi(140)$ &$\rho(775)$  &$\eta(548)$ &$\omega(782)$ & $K(495)$ & $K^*(892)$ \\ \hline
  $-$ &93 &800 &611 &705 &326 &965 \\ \hline
    & $\eta'(958)$ &$\phi(1019)$  &$D^0(1865)$ &$D^*(2007)$   \\ \hline
  $-$ &914 &1056 &1842 &2043 & & \\ \hline
\end{tabular}
\end{table}

From Table \ref{SBM}, we can see that the masses of $P$-wave $\Omega_c$'s are higher than 3200 MeV
although the mass of $P$-wave nucleon is close the experimental value (for the set I). The parameters
of set II is used to check the dependence of the results on the model parameters. The results show that
the $P$-wave baryons have rather large masses, comparing with the experimental data. So it is still
difficult to have a good description of the negative parity states of baryons in the quark model.
In the following, we use set I parameters to study the 5-quark states.

The wavefunctions for the system are constructed just as the way in Ref.~\cite{YG}. Here only the wavefunctions
of each degree of freedom for five-quark system and parts of the sub-clusters of three-quark and quark-antiquark are listed.
One need to notice that there are many different ways to construct the wave-functions of the system. However,
it makes no difference by choosing any one configuration if all the possible coupling are considered.

For the $\Omega_c^0$ with quark content $sscq\bar{q},~q=u,d,s$ in flavor SU(3) case, there are two types of separation,
one is $(qss)\bar{q}c)$ and the other is $(ssc)\bar{q}q$.
The flavor wavefunctions for the sub-clusters constructed are shown below.
\begin{eqnarray}
&& B_{00}^1 =ssc, ~~~~B_{00}^2 =sss,  \nonumber \\
&& B_{\frac12,\frac12}^1 = \frac{1}{\sqrt{6}}(sus+uss-2ssu), \nonumber \\
&& B_{\frac12,-\frac12}^1 = \frac{1}{\sqrt{6}}(sds+dss-2ssd), \nonumber \\
&&B_{\frac12,\frac12}^2 = \frac{1}{\sqrt{2}}(us-su)s, \nonumber \\
&& B_{\frac12,-\frac12}^2 = \frac{1}{\sqrt{2}}(ds-sd)s, \nonumber \\
&& B_{\frac12,\frac12}^3 =\frac{1}{\sqrt{3}}(ssu+sus+uss),   \\
&&B_{\frac12,-\frac12}^3=\frac{1}{\sqrt{3}}(ssd+sds+dss),  \nonumber \\
&&B_{\frac12,\frac12}^4=\frac{1}{\sqrt{2}}(us+su)c,  \nonumber \\
&&B_{\frac12,-\frac12}^4=\frac{1}{\sqrt{2}}(ds+sd)c,  \nonumber \\
&&B_{\frac12,\frac12}^5=\frac{1}{\sqrt{2}}(us-su)c, \nonumber \\
&&  B_{\frac12,-\frac12}^5=\frac{1}{\sqrt{2}}(ds-sd)c,  \nonumber \\
&&
M^1_{\frac12,\frac12} = \bar{d}c, ~~~~
M^1_{\frac12,-\frac12} = -\bar{u}c, \nonumber \\
&& M^2_{\frac12,\frac12} = \bar{d}s, ~~~~
M^2_{\frac12,-\frac12} = -\bar{u}s, \\
&&M_{00}^1 = \frac{1}{\sqrt{2}}(\bar{u}u+\bar{d}d) ,~~
M_{00}^2 = \bar{s}s,~~M_{00}^3 = \bar{s}c. \nonumber
\end{eqnarray}
The flavor wavefunctions for 5-quark system with isospin $I=0$ are obtained by
the following couplings,
\begin{eqnarray}
\chi^f_1  & = & \sqrt{\frac{1}{2}} (B^1_{\frac12,\frac12} M^1_{\frac12,-\frac12}
 - B^1_{\frac12,-\frac12} M^1_{\frac12,\frac12}), \nonumber \\
\chi^f_2  & = & \sqrt{\frac{1}{2}} (B^2_{\frac12,\frac12} M^1_{\frac12,-\frac12}
 - B^2_{\frac12,-\frac12} M^1_{\frac12,\frac12}), \nonumber \\
\chi^f_3  & = & \sqrt{\frac{1}{2}} (B^3_{\frac12,\frac12} M^1_{\frac12,-\frac12}
 - B^3_{\frac12,-\frac12} M^1_{\frac12,\frac12}), \\
\chi^f_4  & = & \sqrt{\frac{1}{2}} (B^4_{\frac12,\frac12} M^2_{\frac12,-\frac12}
 - B^4_{\frac12,-\frac12} M^2_{\frac12,\frac12}), \nonumber \\
\chi^f_5  & = & \sqrt{\frac{1}{2}} (B^5_{\frac12,\frac12} M^2_{\frac12,-\frac12}
 - B^5_{\frac12,-\frac12} M^2_{\frac12,\frac12}), \nonumber \\
\chi^f_6  & = & B_{00}^1 M_{00}^1, ~~\chi^f_7  = B_{00}^1 M_{00}^2,
~~\chi^f_8  = B_{00}^2 M_{00}^3.  \nonumber
\end{eqnarray}

In a similar way, the spin and color wavefunctions for 5-quark system can be constructed,
which are the same as the expressions of Ref.~\cite{YG}. Here we only give the expressions
of 5-quark system, the wavefunctions for the sub-clusters can be found in Ref.~\cite{YG}.
\begin{eqnarray}
& & \chi_{\frac12,\frac12}^{\sigma 1}(5) = \sqrt{\frac{1}{6}}
\chi_{\frac32,-\frac12}^{\sigma}(3) \chi_{11}^{\sigma}
-\sqrt{\frac{1}{3}} \chi_{\frac32,\frac12}^{\sigma}(3) \chi_{10}^{\sigma}
\nonumber \\
&& ~~~~~~~~~~~~+\sqrt{\frac{1}{2}} \chi_{\frac32,\frac32}^{\sigma}(3) \chi_{1-1}^{\sigma} \nonumber \\
& & \chi_{\frac12,\frac12}^{\sigma 2}(5) = \sqrt{\frac{1}{3}}
\chi_{\frac12,\frac12}^{\sigma 1}(3) \chi_{10}^{\sigma}\rangle
-\sqrt{\frac{2}{3}} \chi_{\frac12,-\frac12}^{\sigma 1}(3) \chi_{11}^{\sigma}  \nonumber \\
& & \chi_{\frac12,\frac12}^{\sigma 3}(5) = \sqrt{\frac{1}{3}}
\chi_{\frac12,\frac12}^{\sigma 2}(3) \chi_{10}^{\sigma}
-\sqrt{\frac{2}{3}} \chi_{\frac12,-\frac12}^{\sigma 2}(3) \chi_{11}^{\sigma}  \nonumber \\
& & \chi_{\frac12,\frac12}^{\sigma 4}(5) = \chi_{\frac12,\frac12}^{\sigma 1}(3)
  \chi_{00}^{\sigma}   \nonumber \\
& & \chi_{\frac12,\frac12}^{\sigma 5}(5) = \chi_{\frac12,\frac12}^{\sigma 2}(3)
  \chi_{00}^{\sigma}   \\
& & \chi_{\frac32,\frac32}^{\sigma 1}(5) = \sqrt{\frac{3}{5}}
\chi_{\frac32,\frac32}^{\sigma}(3)\rangle
  \chi_{10}^{\sigma} -\sqrt{\frac{2}{5}} \chi_{\frac32,\frac12}^{\sigma}(3)
  \chi_{11}^{\sigma} \nonumber \\
& & \chi_{\frac32,\frac32}^{\sigma 2}(5) = \chi_{\frac32,\frac32}^{\sigma}(3)
  \chi_{00}^{\sigma}  \nonumber \\
& & \chi_{\frac32,\frac32}^{\sigma 3}(5) = \chi_{\frac12,\frac12}^{\sigma 1}(3)
  \chi_{11}^{\sigma}  \nonumber \\
& & \chi_{\frac32,\frac32}^{\sigma 4}(5) = \chi_{\frac12,\frac12}^{\sigma 2}(3)
  \chi_{11}^{\sigma}  \nonumber \\
& & \chi_{\frac52,\frac52}^{\sigma 1}(5) = \chi_{\frac32,\frac32}^{\sigma}(3)
  \chi_{11}^{\sigma}  \nonumber
\end{eqnarray}
\begin{eqnarray}
\chi^c_1 & = & \frac{1}{\sqrt{18}}(rgb-rbg+gbr-grb+brg-bgr) \nonumber \\
  & & ~~~~~~(\bar r r+\bar gg+\bar bb), \\
\chi^{c}_k & = & \frac{1}{\sqrt{8}}(\chi^k_{3,1}\chi_{2,8}-\chi^k_{3,2}\chi_{2,7}-\chi^k_{3,3}\chi_{2,6}
+\chi^k_{3,4}\chi_{2,5} \nonumber \\
  & &    +\chi^k_{3,5}\chi_{2,4}-\chi^k_{3,6}\chi_{2,3}-\chi^k_{3,7}\chi_{2,2}+\chi^k_{3,8}\chi_{2,2}),
\end{eqnarray}
with $k=2,3$. For the color part, both the color singlet channels ($k=1$) and the hidden color channels
($k=2,3$), are considered here to have an economic way to describe multi-quark system~\cite{YG}.

For the orbital wavefunctions, there are four relative motions for 5-body system. In the present work, the orbital
wavefunctions for each relative motion of the system are determined by the dynamics of the system, The orbital
wavefunctions for this purpose is written as follows,
\[
\psi_{LM_L}=\left[ \left[ \left[
  \phi_{n_1l_1}(\mbox{\boldmath $\rho$})\phi_{n_2l_2}(\mbox{\boldmath $\lambda$})\right]_{l}
  \phi_{n_3l_3}(\mbox{\boldmath $r$}) \right]_{l^{\prime}}
  \phi_{n_4l_4}(\mbox{\boldmath $R$}) \right]_{LM_L}
\]
where the Jacobi coordinates are defined as,
\begin{eqnarray}
{\mbox{\boldmath $\rho$}} & = & {\mbox{\boldmath $x$}}_1-{\mbox{\boldmath $x$}}_2, \nonumber \\
{\mbox{\boldmath $\lambda$}} & = & {\mbox{\boldmath $x$}}_3
 -(\frac{{m_1\mbox{\boldmath $x$}}_1+{m_2\mbox{\boldmath $x$}}_2}{m_1+m_2}),  \\
{\mbox{\boldmath $r$}} & = & {\mbox{\boldmath $x$}}_4-{\mbox{\boldmath $x$}}_5, \nonumber \\
{\mbox{\boldmath $R$}} & = & \left(\frac{{m_1\mbox{\boldmath $x$}}_1+{m_2\mbox{\boldmath $x$}}_2
  +{m_3\mbox{\boldmath $x$}}_3}{m_1+m_2+m_3}\right)
  -\left(\frac{{m_4\mbox{\boldmath $x$}}_4+{m_5\mbox{\boldmath $x$}}_5}{m_4+m_5}\right). \nonumber
\end{eqnarray}
To find the orbital wavefunctions, the gaussian expansion method (GEM) is employed, i.e.,
every $\phi$ is expanded by gaussians with various sizes~\cite{GEM}
\begin{equation}
 \phi_{nlm}(\mbox{\boldmath $r$})=\sum_{n=1}^{n_{max}} c_n N_{nl}r^le^{-(r/r_n)^2}Y_{lm}(\hat{\mbox{\boldmath $r$}}),
\end{equation}
where $N_{nl}$ is the normalization constant,
\begin{equation}
 N_{nl}=\left[\frac{2^{l+2}(2\nu_n)^{l+\frac{3}{2}}}{\sqrt \pi(2l+1)}\right]^{\frac{1}{2}}.
\end{equation}
The size parameters of gaussians $r_n$ are taken as the geometric progression numbers
\begin{equation}
 r_n=r_{1}a^{n-1}.
\end{equation}
$c_n$ is the variational parameters, which is determined by the dynamics of the system.

Finally, the complete channel wave function for the 5-quark system is written as
\begin{equation}
 \Psi_{JM,i,j,k,n}={\cal A} \left[ \left[
   \chi^{\sigma_i}_{S}(5) \psi_{L}\right]_{JM_J}
   \chi^{f}_j \chi^{c}_k \right] \label{wf}
\end{equation}
where ${\cal A}$ is the antisymmetry operator of the system. In the flavor SU(3) case, it has six terms for
the system with three identical particles and it can be reduced to three terms, as follows, due to the symmetry
between first two particles has been considered when constructing the wavefunctions of the 3-quark clusters.
For the two types of separations, 1-$(uss)(\bar{u}c)+(dss)(\bar{d}c),(sss)(\bar{s}c)$,
2-$(ssc)(\bar{u}u+\bar{d}d),(ssc)(\bar{s}s)$, we have the following antisymmetric operators,
\begin{eqnarray}
 {\cal A}_1 & = & 1-(13)-(23), \\
 {\cal A}_2 & = & 1-(15)-(25).
\end{eqnarray}

The eigen-energy of the system is obtained by solving the following eigen-equation
\begin{equation}
H\Psi_{JM}=E\Psi_{JM},
\end{equation}
by using variational principle. The eigen functions $\Psi_{JM}$ are the linear combination of the
above channel wavefunctions Eq.(\ref{wf}).

In evaluating the matrix elements of hamiltonian, the calculation is rather complicated, if the
orbital angular momenta of relative motions of system are not all zero. Here a useful method named
the infinitesimally-shifted gaussian are used~\cite{GEM}. In this method, the spherical harmonic
function is absorbed into the shifted gaussians,
\[
  \phi_{nlm}(\mbox{\boldmath $r$})=N_{nl}\lim_{\varepsilon\to 0}\frac{1}{(\nu \varepsilon)^l}
  \sum_{k=1}^{k_{max}}C_{lm,k}e^{{-\nu_n (\mbox{\boldmath $r$}-\varepsilon \mbox{\boldmath $D$}_{lm,k})}^2},
\]
the calculation becomes easy with no tedious angular-momentum algebra required.

\section{Results and discussions}
In the present calculation, we are interested in the low-lying states of $ussc\bar{u}$, $dssc\bar{d}$
pentaquark system, so all the orbital angular momenta are set to 0. Then the parity of
five-quark system with one antiquark is negative. In this way, the total angular momentum $J$ can
take values 1/2, 3/2 and 5/2. The possible channels under the consideration are listed in
Tables \ref{channel1}-\ref{channel5}.
\begin{table}[ht]
\caption{The channels with $IJ^P=0\frac{1}{2}^-$.  \label{channel1}}
\begin{tabular}{cccccc} \hline
  index & $\chi_{1/2}^{\sigma_i}$ & $\chi_j^f$ & $\chi_k^c$ & physical channel \\ \hline
     1  & $i=1$   & $j=3$   & $k=1$   & $\Xi^* \bar{D}^*$ \\
     2  & $i=1$   & $j=3$   & $k=3$   &   \\
     3  & $i=1$   & $j=4$   & $k=1$   & $\Xi_c^* \bar{K}^*$ \\
     4  & $i=1$   & $j=4,5$ & $k=2,3$ &   \\
     5  & $i=1$   & $j=6$   & $k=1$   & $\Omega_c^* \omega$ \\
     6  & $i=1$   & $j=6$   & $k=3$   &   \\
     7  & $i=2,3$ & $j=1,2$ & $k=1$   & $\Xi \bar{D}^*$ \\
     8  & $i=2,3$ & $j=1,2$ & $k=2,3$ &   \\
     9  & $i=2,3$ & $j=4,5$ & $k=1$   & $\Xi_c \bar{K}^*$ \\
    10  & $i=2,3$ & $j=4,5$ & $k=2,3$ &   \\
    11  & $i=2$   & $j=6$   & $k=1$   & $\Omega_c \omega$ \\
    12  & $i=2,3$ & $j=6$   & $k=2,3$ &   \\
    13  & ~~$i=4,5$~~ & ~~$j=1,2$~~ & $k=1$   & $\Xi \bar{D}$ \\
    14  & $i=4,5$ & $j=1,2$ & ~~$k=2,3$~~ &   \\
    15  & ~~$i=4,5$~~ & ~~$j=4,5$~~ & $k=1$   & $\Xi_c \bar{K}$ \\
    16  & $i=4,5$ & $j=4,5$ & ~~$k=2,3$~~ &   \\
    17  & $i=4$   & $j=6$   & $k=1$   & $\Omega_c \eta$ \\
    18  & $i=4,5$ & $j=6$   & $k=2,3$ &   \\
 \hline
\end{tabular}
\end{table}

\begin{table}[ht]
\caption{The channels with $IJ^P=0\frac{3}{2}^-$.  \label{channel3}}
\begin{tabular}{cccccc} \hline
  index & $\chi_{3/2}^{\sigma_i}$ & $\chi_j^f$ & $\chi_k^c$ & physical channel \\ \hline
     1  & $i=1$   & $j=3$   & $k=1$   & $\Xi^* \bar{D}^*$ \\
     2  & $i=1$   & $j=3$   & $k=3$   &   \\
     3  & $i=1$   & $j=4$   & $k=1$   & $\Xi_c^* \bar{K}^*$ \\
     4  & $i=1$   & $j=4,5$ & $k=2,3$ &   \\
     5  & $i=1$   & $j=6$   & $k=1$   & $\Omega_c^* \omega$ \\
     6  & $i=1$   & $j=6$   & $k=3$   &   \\
     7  & $i=2$   & $j=3$   & $k=1$   & $\Xi^* \bar{D}$ \\
     8  & $i=2$   & $j=3$   & $k=3$   &   \\
     9  & $i=2$   & $j=4$   & $k=1$   & $\Xi_c^* \bar{K}$ \\
    10  & $i=2$   & $j=4,5$ & $k=2,3$ &   \\
    11  & $i=2$   & $j=6$   & $k=1$   & $\Omega_c^* \eta$ \\
    12  & $i=2$   & $j=6$   & $k=3$ &   \\
    13  & ~~$i=3,4$~~ & ~~$j=1,2$~~ & $k=1$   & $\Xi \bar{D}^*$ \\
    14  & $i=3,4$ & $j=1,2$ & ~~$k=2,3$~~ &   \\
    15  & ~~$i=3,4$~~ & ~~$j=4,5$~~ & $k=1$   & $\Xi_c \bar{K}^*$ \\
    16  & $i=3,4$ & $j=4,5$ & ~~$k=2,3$~~ &   \\
    17  & $i=3$ & $j=6$   & $k=1$   & $\Omega_c \omega$ \\
    18  & $i=3,4$ & $j=6$   & $k=2,3$ &   \\
 \hline
\end{tabular}

\caption{The channels with $IJ^P=0\frac{5}{2}^-$.  \label{channel5}}
\begin{tabular}{cccccc} \hline
  index & $\chi_{5/2}^{\sigma_i}$ & $\chi_j^f$ & $\chi_k^c$ & physical channel \\ \hline
     1  & $i=1$   & $j=3$   & $k=1$   & $\Xi^* \bar{D}^*$ \\
     2  & $i=1$   & $j=3$   & $k=3$   &   \\
     3  & $i=1$   & $j=4$   & $k=1$   & $\Xi_c^* \bar{K}^*$ \\
     4  & ~~$i=1$~~   & ~~$j=4,5$~~ & ~~$k=2,3$~~ &   \\
     5  & $i=1$   & $j=6$   & $k=1$   & $\Omega_c^* \omega$ \\
     6  & $i=1$   & $j=6$   & $k=3$   &   \\
 \hline
\end{tabular}
\end{table}

\begin{table}[hbt]
\caption{The lowest eigen-energies of the $udc{\bar{c}}u$ system with $J^P=\frac12^-$ (unit: MeV).
 The percentages of color-singlet (S) and hidden-color (H) channels are also given. \label{Gresult1}}
\begin{tabular}{lccccc} \hline
   Channel   & ~~~~$E$~~~~  & $E_{th}^{Theo}$   & $E_B$ &
    $E_{th}^{Exp}$   &  $E'$ \\ \hline
    1   & 3526 & 3531 & $-5$ & 3539($\Xi^*\bar{D}^*$)  & 3534 \\
    2   & 4016 &      &      &   &  \\
    1+2 & 3525 &      & $-6$ &   & 3533 \\
        &      &  \multicolumn{4}{l}{percentage(S;H): 99.8\%; 0.2\%}  \\ \hline
    3   & 3566 & 3568 & $-2$ & 3537($\Xi_c^* \bar{K}^*$)       & 3535 \\
    4   & 3616 &      &      &        &  \\
    3+4 & 3564 &      & $-4$ &    &  3533 \\
        &      &  \multicolumn{4}{l}{percentage(S;H): 96.3\%; 3.7\%}  \\ \hline
    5   & 3453 & 3453 &   0   & 3548($\Omega_c^* \omega$)  & 3453 \\
    6   & 3404 &  &     &        &  \\
    5+6 & 3402 &      & $-51$ &    & 3497 \\
        &      &  \multicolumn{4}{l}{percentage(S;H): 0.2\%; 99.8\%}  \\ \hline
    7   & 3374 & 3405 & $-31$    & 3322($\Xi \bar{D}^*$)    & 3291 \\
    8   & 3672 &  &     &     &  \\
    7+8 & 3373 &      & $-32$   &    & 3290 \\
        &      &  \multicolumn{4}{l}{percentage(S;H): 99.8\%; 0.2\%}  \\ \hline
    9   & 3495 & 3506 & $-11$    & 3359($\Xi_c \bar{K}^*$)       & 3348 \\
    10  & 3613 &  &     &        &  \\
    9+10 & 3472 &     & $-34$   &   &  3325 \\
        &       &  \multicolumn{4}{l}{percentage(S;H): 85.2\%; 14.8\%}  \\ \hline
    11  & 3380 & 3380  & 0    & 3477($\Omega_c \omega$)       & 3477 \\
    12  & 3608 &   &     &        &  \\
    11+12 & 3380 &  \multicolumn{4}{l}{ }  \\ \hline
    13  & 3175 & 3204 & $-29$    & 3185($\Xi\bar{D}$)    & 3156 \\
    14  & 3811 &  &     &     &  \\
    13+14  & 3175 &     & $-29$   &    & 3156 \\
        &      &   \multicolumn{4}{l}{percentage(S;H): 100.0\%; 0.0\%}  \\ \hline
    15  & 2867 & 2867 & 0    & 2961($\Xi_c \bar{K}$)       & 2961 \\
    16  & 3807 &  &     &        &  \\
    15+16  & 2855 &     & $-12$    &   &   2949   \\
          &       &  \multicolumn{4}{l}{percentage(S;H): 96.7\%; 3.3\%}  \\ \hline
    17  & 3286 & 3286 & 0    & 3243($\Omega_c \eta$)       & 3243 \\
    18  & 3828 &  &     &        &  \\
    17+18 & 3286 &  \multicolumn{4}{l}{ }  \\ \hline
    mixed (singlet) & 2771 & 2867 &  $-96$  & 2961($\Xi_c \bar{K}$) & 2865  \\
    mixed (full)    & 2675 & 2867 & $-192$  & 2961($\Xi_c \bar{K}$) & 2769 \\  \hline
\end{tabular}
\end{table}

First, the single channel calculations are performed. The eigen-energies of each states with
different quantum numbers are shown in Tables \ref{Gresult1}-\ref{Gresult3}, where the
eigen-energies of the states are shown in column 2, along with the theoretical thresholds
in column 3 and experimental thresholds in column 5, column 4 gives the binding energies,
the difference between the eigen-energies and the theoretical thresholds,
$E_B=E-E_{th}^{Theo}$. The corrected energies of the states (column 6), which are obtained
by taking the sum of experimental thresholds and the binding energies. Namely, and $E'=E_B+E_{th}^{Exp}$.

Secondly, the three types of channel coupling calculations are performed. The first is the
channel coupling between color-singlet and hidden-color channels with the same flavor-spin
structures. The second is the coupling among all color-singlet channels with different flavor-spin
structures and the last is the full coupling, including all channels for given $J^P$.
Table \ref{Distance} gives the spacial configurations of the states by calculating the
distances between any two quarks or quark and antiquark in the full channel coupling calculation.

In the following we analyze the results in detail.

(a) $J^P=\frac{1}{2}^-$: The single channel calculations show that there are weak attractions
for the most channels, the exceptions are $\Omega_c \eta, \Omega_c \omega, \Omega_c^* \omega$
and $\Xi_c \bar{K}$. The coupling to hidden-color channels helps a little, increasing the attraction
a few MeVs and pushing $\Omega_c^* \omega$ and $\Xi_c \bar{K}$ below the corresponding thresholds. So
the resonances can be formed. Most of the states have higher masses compared with that of the five new
excited states of $\Omega_c$. For $\Xi\bar{D}$, the second lowest state, it has the energy 3156 MeV,
which is close to the highest $\Omega_c$, 3119 MeV. The lowest state $\Xi_c\bar{K}$ has the energy
2949 MeV with the help of hidden-color channel coupling, which is a little smaller than the mass of the
lowest excited state of $\Omega_c$, 3000 MeV.
\begin{table}[htb]
\caption{The eigen-energies of full channel-coupling calculation below 3.2 GeV with $IJ^P=0\frac{1}{2}^-$.
(unit: MeV).  \label{cc12}}
\begin{tabular}{ccccccc} \hline
  index        &   1  &   2  &  3   &  4   &  5   &  6       \\ \hline
  ~~$E$~~      & ~2675~ & ~2867~ & ~2873~ & ~2882~ & ~2901~ & ~2937~  \\
  $E^{\prime}$ & 2769 & 2961 & 2967 & 2976 & 2995 & 3031  \\ \hline
\end{tabular}

\caption{The lowest eigen-energies of the $udc{\bar{c}}u$ system with $\frac32^-$(unit: MeV).
  \label{Gresult2}}
\begin{tabular}{lccccc} \hline
   Channel   & ~~~~$E$~~~~  & $E_{th}^{Theo}$   & $E_B$ &
    $E_{th}^{Exp}$   &  $E'$ \\ \hline
    1    & 3521 & 3531 & $-10$  & 3539($\Xi^* \bar{D}^*$)   & 3529 \\
    2    & 4026 &      &        &                           &  \\
    1+2  & 3521 &      & $-10$  &                           & 3529 \\
         &      &  \multicolumn{4}{l}{percentage(S;H): 100.0\%; 0.0\%}  \\ \hline
    3    & 3565 & 3568 & $-3$   & 3537($\Xi_c^* \bar{K}^*$) & 3534 \\
    4    & 3617 &      &        &                           &  \\
    3+4  & 3562 &      & $-6$   &                           & 3531 \\
         &      &  \multicolumn{4}{l}{percentage(S;H): 94.0\%; 6.0\%}  \\ \hline
    5    & 3453 & 3453 & 0      & 3548($\Omega_c^* \omega$) & 3548 \\
    6    & 3477 &      &        &                           &  \\
    5+6  & 3453 &  \multicolumn{4}{l}{ }  \\ \hline
    7    & 3309 & 3330 & -21    & 3397($\Xi^*\bar{D}$)  & 3376 \\
    8    & 4145 &      &        &                       &  \\
    7+8  & 3309 &      & $-21$  &                       & 3376 \\
         &      &  \multicolumn{4}{l}{percentage(S;H): 100.0\%; 0.0\%}  \\ \hline
    9    & 2929 & 2929 & 0      & 3139($\Xi_c^* \bar{K}$) & 3139 \\
    10   & 3782 &      &        &                         &  \\
    9+10 & 2728 &      & $-1$   &                         & 3138 \\
         &      &  \multicolumn{4}{l}{percentage(S;H): 99.7\%; 0.3\%}  \\ \hline
    11   & 3359 & 3359 & 0      & 3314($\Omega_c^* \eta$) & 3314 \\
    12   & 3763 &      &        &                         &  \\
    11+12 & 3359 & \multicolumn{4}{l}{ }  \\ \hline
    13   & 3388 & 3405 & $-17$  & 3322($\Xi\bar{D}^*$)    & 3305 \\
    14   & 3705 &      &        &                         &  \\
    13+14 & 3388 &     & $-17$  &                         & 3305 \\
          &      &  \multicolumn{4}{l}{percentage(S;H): 100.0\%; 0.0\%}  \\ \hline
    15   & 3506 & 3506 & 0      & 3359($\Xi_c \bar{K}^*$) & 3359 \\
    16   & 3656 &      &        &                         &  \\
    15+16 & 3348 &     & $-158$ &                         & 3201 \\
         &      &  \multicolumn{4}{l}{percentage(S;H): 57.7\%; 42.3\%}  \\ \hline
    17   & 3380 & 3380 & 0    & 3477($\Omega_c \omega$)       & 3477 \\
    18   & 3588 &      &      &                               &  \\
    17+18 & 3380 & \multicolumn{4}{l}{ } \\  \hline
    mixed (singlet) & 2928  & 2929  &  $-1$ & 3139($\Xi_c^* \bar{K}$) & 3138 \\
     mixed (full)   & 2857  & 2929  & $-72$ & 3139($\Xi_c^* \bar{K}$) & 3067 \\  \hline
\end{tabular}
\end{table}

The situation changes a lot after coupling all the color-singlet channels, the lowest energy
we obtained is 2865 MeV. And the full channel-coupling calculation decreases the lowest energy
further to 2769 MeV. Table \ref{cc12} shows the six lowest eigen-energies in the full-channel
calculation. $E^{\prime}$ denotes the corrected energy,
$$E^{\prime}=E^{Exp}_{th}(\Xi_c \bar{K})-E^{Theo}_{th}(\Xi_c \bar{K}) +E. $$
In this way, there are many $\Omega_c$ pentaquark states with $IJ^P=0\frac{1}{2}^-$ in the
quark model. To assign these states to the excited $\Omega_c$ states announced by LHCb, further work
is needed. The problem has to be solved is how to correct the eigen-energies from the full
channel-coupling calculation.

One interesting state is $\Omega_c^* \omega$, the hidden-color channel has lower energy than the
colorless one. It is a possible good resonance because of its color structure, although it has
a rather high energy, 3497 MeV.

(b) $J^P=\frac{3}{2}^-$: We have similar results with that of $J^P=\frac{1}{2}^-$. Four channels,
$\Omega_c^* \omega, \Xi_c^* \bar{K}, \Xi_c \bar{K}^*$ and $\Omega_c \omega$, have no attraction
in single channel calculations. and the hidden-color channel-coupling induces a very weak attraction
for $\Xi_c^* \bar{K}$. But, it introduces a large attraction for $\Xi_c \bar{K}^*$, $-158$ MeV,
a good candidate of color structure resonance to be confirmed.

All color-singlet channel-coupling calculation gives a very weak bound state with energy 3138 MeV
after correction. The full channel-coupling lowered the energy further to 3067 MeV. Table \ref{cc32}
shows the four lowest eigen-energies in the full-channel coupling calculation. After correction,
their energies are below 3.2 GeV.

(c) $J^P=\frac{5}{2}^-$: Only one channel, $\Xi^*\bar{D}^*$, has attractive in the single channel
calculation. Coupling to the hidden-color channels, an additional channel, $\Xi_c^* \bar{K}^*$,
is induced out an attraction. Channel-couplings, color-singlet and full, do not produce any
bound state. The $D$-wave $\Xi$-$\bar{D}$ and/or $\Xi_c$-$\bar{K}$ scattering phase shift calculation is needed
to check that the resonances, $\Xi^*\bar{D}^*$ and $\Xi_c^* \bar{K}^*$, can survive or not after
the coupling.

\begin{table}[htb]
\caption{The eigen-energies of full channel-coupling calculation below 3.2 GeV with $IJ^P=0\frac{3}{2}^-$.
(unit: MeV).  \label{cc32}}
\begin{tabular}{cccccccc} \hline
  index        &   1  &   2  &  3   &  4       \\ \hline
  ~~$E$~~      & ~2857~ & ~2931~ & ~2940~ & ~2956~  \\
  $E^{\prime}$ & 3067 & 3141 & 3150 & 3166    \\ \hline
\end{tabular}

\caption{The lowest eigen-energies of the $ssc{\bar{u}}u$+$ssc{\bar{d}}d$ system with $\frac52^-$(unit: MeV).
  \label{Gresult3}}
\begin{tabular}{lccccc} \hline
   Channel   & ~~~~$E$~~~~  & $E_{th}^{Theo}$   & $E_B$ &
    $E_{th}^{Exp}$   &  $E'$ \\ \hline
    1   & 3508 & 3531 & $-23$ & 3539($\Xi^*\bar{D}^*$) & 3516 \\
    2   & 4042 &      &       &                        & \\
    1+2 & 3507 &      & $-24$ &                        & 3515 \\
        &      &  \multicolumn{4}{l}{percentage(S;H): 99.8\%; 0.2\%}  \\ \hline
    3   & 3568 & 3568 & 0     & 3537($\Xi_c^* \bar{K}^*$) & 3537 \\
    4   & 3646 &      &       &                           & \\
    3+4 & 3532 &      & $-36$ &                           & 3501 \\
               &      &  \multicolumn{4}{l}{percentage(S;H): 80.0\%; 20.0\%}  \\ \hline
    5   & 3453 & 3453 & 0     & 3548($\Omega_c^*\omega$)  & 3548 \\
    6   & 3563 &      &       &                           & \\
    5+6 & 3453 &      &       &                           &  \\ \hline
    mixed (singlet)  & 3453 & &    &                      &  \\
    mixed (full)     & 3453 & &    &                      & \\  \hline
\end{tabular}
\end{table}

\begin{table}[htb]
\caption{Distances between quarks, $q$ is for $u,d$ quark and $Q$ is for $c$ quark (unit: fm).  \label{Distance}}
\begin{tabular}{cccccc} \hline
  $J^{P}$  & ~~~~Channel~~~~   & ~~$r_{qq}$~~ & ~~$r_{qQ}$~~ & ~~$r_{q\bar{q}}$~~ & ~~$r_{Q\bar{q}}$~~  \\ \hline
 ${\frac{1}{2}}^{-}$ & $\Omega_c^0$(2769)  & 1.3  & 1.1  & 1.4 & 1.2 \\
 \hline
 ${\frac{3}{2}}^{-}$ & $\Omega_c^0$(3067)    & 1.4  & 1.1 & 1.2 & 1.4 \\
\hline
\end{tabular}
\end{table}

Table \ref{Distance} gives the distances between quarks for two states, $\Omega_c^0$(2769)
and $\Omega_c^0$(3067). All the quark-pairs have similar distances and all are smaller than 1.5 fm.
So these two states are compact ones.

\section{Summary}
In the framework of the chiral quark model, the 5-quark systems with quark contents
$sscu\bar{u}$, $sscd\bar{d}$ are investigated by means of Gaussian expansion method.
The calculation shows that there are several resonance states for
$I(J^P)=0({\frac{1}{2}}^-)$, $0({\frac{3}{2}}^-)$ below 3.2 GeV. $\Xi\bar{D}$, $\Xi_c\bar{K}$
and $\Xi_c^*\bar{K}$ are possible the candidates of the newly announced excited states of $\Omega_c^{0}$
by LHCb Collaboration. In the present calculation, the masses of the lowest states with quantum numbers 
$IJ^P=0\frac12^{-}$ and $IJ^P=0\frac32^{-}$ are 2769 MeV and 3067 MeV, respectively. And the distances 
between quark pairs suggest these two states are compact states or 
pentaquark structures. It manifests the effects of hidden-color channels. So it is interesting to identify
the states experimentally. In this work, in fact we cannot identify the excited states of $\Omega_c^{0}$
reported by LHCb Collaboration with the pentaquarks we calculated. We want stress that the $P$-wave
$q^3$ baryon will mix strongly with the $S$-wave pentaquark. The unquenched quark model, including the
high Fock components, study of $\Omega_c$ is needed to clarify the situation.

In the present calculation, the internal structures of the sub-clusters are not fixed,
the structure of a 5-quark system is determined by the dynamics of the system, because all
the possible coupling are included except the high orbital angular momentum.
The further work of considering the high orbital angular momenta along with the spin-orbit and
tensor interactions is expected.

Pentaquark involves two subcluster, $q^3$ and $q\bar{q}$. If the two subclusters are colorless,
they are corresponding to baryon and meson. To describe baryon and meson simultaneously in quark
model with one set of parameters is still difficult. It is main source of the uncertainty of 
the model calculation of pentaquark. Unquenched quark model may be a solution for the 
unified description of baryon and meson, since the $q\bar{q}$ cluster is always involved. 
 
Multiquark states are ideal place to develop the quark model. Because the model approach is a 
phenomenogical one, its development depends on the accumulated experimental data.
We hope that the model description of the multiquark states will be improved with the accumulation 
of the experimental data on multiquark state,

\section*{Acknowledgments}
The work is supported partly by the National Natural Science Foundation of China under Grant
Nos. 11535005, 11175088, and 11205091.

\end{document}